\documentclass[aps,prl,twocolumn,superscriptaddress,showpacs,floatfix]{revtex4-2}
\usepackage{graphicx,bm,epsfig,textcomp,color,dcolumn,setspace,array,calrsfs,mathrsfs,amsmath,amssymb,mathptmx,gensymb,booktabs,url,bibentry,natbib,subfigure,float,braket,lipsum,xcolor,relsize}
\usepackage[T1]{fontenc}
\usepackage[mathscr]{eucal}
\usepackage[bookmarks=true,colorlinks,linkcolor=blue,urlcolor=blue,citecolor=blue,plainpages=false,pdfpagelabels,final,breaklinks=true]{hyperref}
\DeclareMathAlphabet{\mathcal}{OMS}{cmsy}{m}{n}

\begin{document}
\title{Lossless Spin-Orbit Torque in Antiferromagnetic Topological Insulator MnBi$_2$Te$_4$}
\author{Junyu Tang}
\affiliation{Department of Physics and Astronomy, University of California, Riverside, California 92521, USA}
\author{Ran Cheng}
\affiliation{Department of Electrical and Computer Engineering, University of California, Riverside, California 92521, USA}
\affiliation{Department of Physics and Astronomy, University of California, Riverside, California 92521, USA}

\begin{abstract}
We formulate and quantify the spin-orbit torque (SOT) in intrinsic antiferromagnetic topological insulator $\rm MnBi_2Te_4$ of a few septuple-layer thick in charge-neutral condition, which exhibits pronounced layer-resolved characteristics and even-odd contrast. Contrary to traditional current-induced torques, our SOT is not accompanied by Ohm's currents, thus being devoid of Joule heating. We study the SOT-induced magnetic resonances, where in the tri-septuple-layer case we identify a peculiar exchange mode that is blind to microwaves but can be exclusively driven by the predicted SOT. As an inverse effect, the dynamical magnetic moments generate a pure adiabatic current, which occurs concomitantly with the SOT and gives rise to an overall reactance for the $\rm MnBi_2Te_4$, enabling a lossless conversion of electric power into magnetic dynamics.
\end{abstract}

\maketitle

A central theme of modern spintronics has been the quest for efficient electrical control of magnetism and magnetic dynamics~\cite{Roadmap,Manchon_RMP_2019,Hoffmann_Phys.Rev.Appl_2015,RevModPhys.76.323}. In established paradigms, such control is exemplified by an engineered heterostructure consisting of a magnetic material and a spin generator (\textit{e.g.}, a heavy metal) converting currents into spin angular momenta~\cite{L.J.Zhu_PRL_2019,L.Q.Liu_PRL_2017,D.C.Ralph_Nature_2014,K.L.Wang_NatMaterial_2014,L.Q.Liu_Science_2012,STFMR_Liu,miron2011perpendicular,mihai2010SOT}. This archetypal setup suffers from two serious drawbacks: 1) The spatial separation (\textit{i.e.}, the interface) of electric and magnetic components inhibits the spin-transfer process. 2) The driving electric power is largely dissipated via Joule heating.

Exploiting \textit{intrinsic} magnetic topological materials could could enable direct control of magnetic dynamics via electric stimuli in the absence of interfaces. This is because the coexistence of spin-orbit interaction and intrinsic magnetic order allows a single material (monostructure) to drive itself without relying on external spin sources. While monostructural spin-orbit torques (SOTs) have been explored in non-topological magnets~\cite{FGT_Brataas2019,kurebayashi2022magnetism,zhang2021gigantic,CrI3SOT_XueFei,Jungwirth_PRL_2014,Zhou_PhysRevApplied_2018,Jungwirth_PRB_2017,Binghai_PRL_2017,wadley2016electrical} and topological semimetals~\cite{NC_Hanke2017mixed,Jungwirth_PRL_2017}, little is known about the subtle interplay between layer-resolved magnetism and topological electrons. Recently studies identified $\rm MnBi_2 Te_4$ as an intrinsic antiferromagnetic (AFM) topological insulator featuring layer-contrasting magnetic order intertwined with the electronic band topology~\cite{Otrokov_Nature_2019,chen2019intrinsic,XuYong_SciAdv_2019,J.Wang_PRL_2019,Q.K.Xue_CPL_2019,Y.B.Zhang_Science_2020,Y.Y.Wang_NatMat_2020,ovchinnikov2021intertwined,S.Q.Yang_PRX_2021,zhao2021evenodd, OtrokovPRL2019}, opening an ideal testing ground to study the electrical manipulation of layer-resolved magnetism.

In this Letter, we formulate and quantify the electric field induced SOT and its inverse effect, adiabatic charge pumping, in $\rm MnBi_2Te_4$ of a few septuple-layer (SL) thick. The SOT is manifestly SL-resolved and displays an evident even-\textit{vs}-odd contrast of the (total) SL-number. The physical consequences of the SOT are demonstrated by the SOT-induced magnetic resonances, where in the 3-SL case we identify a unique chiral mode that is blind to microwave electromagnetic fields but can be excited only by the SL-resolved SOT.

In contrast to conventional SOTs accompanied by charge currents, the SOT in $\rm MnBi_2Te_4$ does not incur Ohm's conduction as the Fermi level lies in the gap under the charge-neutral condition. In our scenario, the output current only arises from the coherent dynamics of magnetic moments as a reciprocal effect of the SOT, which is a pure \textit{adiabatic} effect that produces no Joule heating. Under the combined action of the SOT and the adiabatic charge pumping, a voltage-driven $\rm MnBi_2Te_4$ acquires an effective reactance, whereby $100\%$ of the input electric power can be converted into magnetic dynamics to overcome the Gilbert damping, achieving an unprecedented high efficiency of electrical manipulation of magnetism using a single material. The unique mechanism we pursue here is fundamentally distinct from the voltage-controlled magnetic anisotropy~\cite{VCMA_2000,VCMA_review_2019}, the multiferroic effects~\cite{multiferroic_2012,multiferroic_Yoshinori_2015}, and the piezoelectric effects~\cite{Strain_2012_APL,strain_review_2015}.

\textit{Formalism.}---We start by constructing the Lagrangian to quantify the dynamics of a semiclassical wavepacket $\ket{W}$ for a Bloch electron with the center-of-mass position $\bm{r}_c=\bra{W}\hat{\bm{r}}\ket{W}$ and momentum $\bm{k}_c=\bra{W}\hat{\bm{k}}\ket{W}$~\cite{Q.Niu_RMP_2010}, whose spin couples the unit magnetization vector $\bm{m}^j$ ($j$ is the SL index) through the exchange interaction. The wavepacket is moving adiabatically in 2D while the SL-dependence constitutes an internal degree of freedom that does not break the adiabatic condition~\cite{R.Cheng_PRB_2012,R.Cheng_PRB_2014}. To simplify our notation, we focus on a single energy band well-separated from all other bands, which is non-degenerate (doubly degenerate) for an odd (even) total number of SLs. The Lagrangian density of such a wavepacket perturbed by electromagnetic fields expressed in vector potential $\mathscr{\bm{A}}$ and scalar potential $\varphi$ can be written as
\begin{align}
\mathcal{L}_{em}=&
    \hbar\dot{\bm{r}}_c\cdot(\bm{k}_c-e \mathscr{\bm{A}})-\epsilon(\bm{k}_c)\nonumber \\ 
    &+\hbar\eta^{\dagger}\left( \mathrm{i}d/dt+\bm{A}^k\cdot\dot{\bm{k}}_c+\bm{A}^{m^j}\cdot\dot{\bm{m}}^j
    \right)\eta\label{eq:Lem}
\end{align}
where $\hbar$ is the reduced Planck constant, $e>0$ is the absolute electron charge, and summations of repeated indices are assumed (here and hereafter). For a non-degenerate band, $\eta=1$. For a degenerate band, $\eta$ becomes a column vector specifying the projection of the wavepacket on each sub-band~\cite{R.Cheng_PRB_2012}: $\ket{W}=\int d\bm{k} w(\bm{k}) e^{\mathrm{i}\bm{k}\cdot \bm{r}} \eta_a \ket{u_{a}}$ where $\ket{u_a}$ is the periodic part of the Bloch wavefunction and the spectral profile function $w(\bm{k})$ satisfies $\int d\bm{k} |w(\bm{k})|^2 \bm{k}= \bm{k}_c$. The interplay between the electron and the SL magnetization is characterized by the Berry connection matrices: $[A_{ab}]^\alpha_\mu\equiv \mathrm{i}\bra{u_a}\partial^\alpha_\mu\ket{u_b}$, where $\partial_\mu^\alpha$ stands for $\partial/\partial k_\mu$ when $\alpha=k$ and $\partial/\partial m^j_\mu$ when $\alpha=m^j$. In Eq.~\eqref{eq:Lem}, the band energy of the wavepacket is $\epsilon(\bm{k}_c)=\bra{W}H\ket{W}$, where $H=H_0(\hbar\bm{k}+e\bm{\mathscr{A}})-e \varphi$ with $H_0(\bm{\mathscr{A}}\rightarrow0)$ being the unperturbed Hamiltonian. The spatial variation of $\bm{m}^j$ in the in-plane directions is ignored, otherwise the real-space Berry connection $\bm{A}^r$ will also be present.

The Lagrangian density for the dynamics of $\bm{m}^j$ is $\mathcal{L}_{m^j}=\hbar S(1-\cos\theta^j)\partial_t \phi^j - \mathcal{H}^j$~\cite{G.Tatara_PhysicaE_2019},
where $\theta^j$ and $\phi^j$ are the spherical angles specifying the direction of $\bm{m}^j$, $S$ is the total spin quantum number (of a unit cell in each SL), and $\mathcal{H}^j$ is the magnetic free energy of $\bm{m}^j$ including the inter-SL exchange coupling and the magnetic anisotropy. Applying the Euler-Lagrangian equation on $\mathcal{L}_{tot}=\mathcal{L}_{em}+\sum_j\mathcal{L}_{m^j}$, followed by an integration of $\bm{k}_c$ over the first Brillouin zone~\cite{SM}, we obtain the Landau-Lifshitz-Gilbert (LLG) equation for $\bm{m}^j$ and the in-plane current density as
\begin{subequations}
\label{:eq:H_and_J}
\begin{align}
\dot{\bm{m}}^j&=\gamma (\bm{H}^{\rm eff}_j+\bm{H}_j^{\rm T})\times \bm{m}^j+\alpha_{\rm G} \bm{m}^j \times \dot{\bm{m}}^j \label{eq:torque},\\
\bm{J}&=\sigma_0 \bm{E}+\sigma_{\rm AH} \hat{\bm{z}}\times \bm{E}+ \bm{J}^{\rm p}, \label{eq:current}
\end{align}
\end{subequations}
where $\sigma_0$ and $\sigma_{\rm AH}$ are the longitudinal and quantum anomalous Hall (QAH) conductivity, $\hat{\bm{z}}$ is the unit vector normal to the plane, $\bm{E}=-\bm{\nabla}\varphi-\partial_t \mathscr{\bm{A}}$ is the applied electric field, $\bm{H}^{\rm eff}_{j}=(-1/\hbar\gamma S) \partial (\sum_i\mathcal{H}^i)/\partial \bm{m}^j$ (with $\gamma$ the gyro-magnetic ratio) is the effective magnetic field acting on $\bm{m}^j$ in the absence of electric stimuli, and $\alpha_{\rm G}$ is the Gilbert damping constant. In Eqs.~\eqref{eq:torque} and~\eqref{eq:current}, $\bm{H}_j^{\rm T}$ is the effective field of the SOT acting on $\bm{m}^j$ and $\bm{J}^{\rm p}$ is the current density generated by adiabatic charge pumping. In the Cartesian coordinates,
\begin{subequations}
\label{eq:main}
\begin{align}
{H}_{j,\mu}^{\rm T}&=\frac{ea_0^2}{\gamma \hbar S}\int\frac{{\rm d}^2k}{(2\pi)^2} f(\bm{k}){\rm{Tr}}(\Omega^{k m^j}_{\nu\mu}) E_\nu, \label{eq:SOT}\\
J^{\rm p}_\mu&=-e\int\frac{{\rm d}^2k}{(2\pi)^2} f(\bm{k}){\rm{Tr}}(\Omega^{m^j k}_{\nu\mu}) \dot{m}^j_\nu, \label{eq:pumping}
\end{align}
\end{subequations}
where $a_0$ is the lattice constant, $f(\bm{k})$ is the Fermi-Dirac distribution function, and $\Omega^{\alpha\beta}_{\mu\nu}=\partial^\alpha_\mu A^\beta_\nu-\partial^\beta_\nu A^\alpha_\mu-\mathrm{i}[A^\alpha_\mu, A^\beta_\nu]$ is the Berry curvature, in which the commutation term drops out for odd-SL samples where bands are non-degenerate~\cite{Onsager}. In the even-SL cases, the bands are doubly degenerate, and the trace in Eq.~\eqref{eq:main} also applies to the degenerate subspace, which results from thermal averaging of the inter-sub-bands transitions embedded in the dynamics of $\eta$~\cite{SM}. At the charge-neutral point with the Fermi level $\varepsilon_F$ in the gap, $\sigma_0$ vanishes while $\sigma_{\rm AH}$ is quantized for odd-SL (zero for even-SL) samples due to the momentum-space Berry curvature $\Omega^{kk}$.

\begin{figure}[t]
\centering
\includegraphics[width=\linewidth]{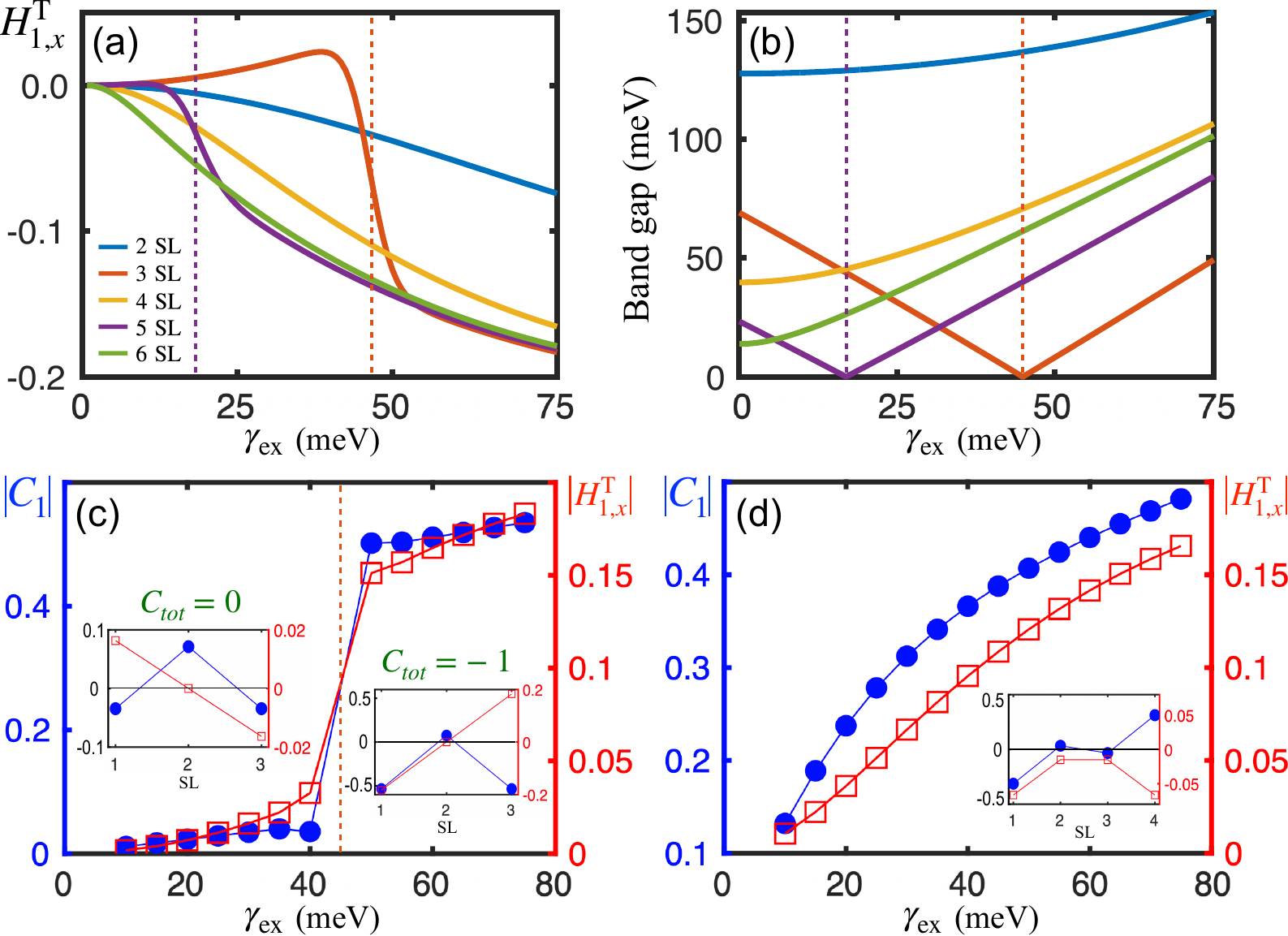}
\caption{(a) Effective field of SOT (in unit of $eEa_0/4\pi^2\hbar\gamma S$) acting on the bottom SL. (b) Band gaps at the $\Gamma$ point as functions of $\gamma_{\rm ex}$ from 2 to 6 SLs. (c) and (d): The bottom-layer Chern number ($C_1$) and the SOT field $H_{1,x}^{\rm T}$ versus $\gamma_{\rm ex}$ for 3-SL and 4-SL $\rm MnBi_2Te_4$, respectively. Insets: layer distributions of the Chern number and the SOT field in the $C_{tot}=0$ phase (at $\gamma_{ex}=30\rm meV$, for both 3-SL and 4-SL) and the $C_{tot}=-1$ phase (at $\gamma_{ex}=75\rm meV$, for 3-SL only).}
\label{fig:SOT&band}
\end{figure}

\textit{Voltage-induced SOT.}---We next calculate the SOT field basing on Eq.~\eqref{eq:SOT} for each SL in a multi-SL $\rm MnBi_2Te_4$. The unperturbed Hamiltonian $H_0(\bm{k})$ comprises the SL-specific $h_j$ and the inter-SL hopping $T_{ij}$ as diagonal and off-diagonal blocks, both of which can be obtained by discretizing the bulk Hamiltonian in the vertical dimension~\cite{J.Wang_PRL_2019,B.Lian_PRL_2020,Y.H.Li_PhysRevResearch_2022}. Under the basis $[\ket{p_{z,\rm Bi}^+,\uparrow},\ket{p_{z,\rm Te}^-,\uparrow},\ket{p_{z,\rm Bi}^+,\downarrow},\ket{p_{z,\rm Te}^-,\downarrow}]^T$,
\begin{subequations}
\begin{align}
 h_j(\bm{k})&=\epsilon(\bm{k})+ d_a(\bm{k})\Gamma_a + \gamma_{\rm ex} \bm{m}^j\cdot\bm{\sigma} \otimes (\tau_0+\delta\tau_3), \label{eq:hl}\\
 T_{ij}&=\sigma_0\otimes(D_1\tau_0+B_1\tau_3) + \mathrm{i}A_1\sigma_z\otimes\tau_1, \label{eq:T0}
\end{align} 
\end{subequations}
where $\epsilon(\bm{k})=C+2D_1+D_2(k_x^2+k_y^2)$, $d_0(\bm{k})=M_0+2B_1+B_2 (k^2_x+k^2_y)$, $d_{1(2)}(\bm{k})=A_2 k_{x(y)}$, $\gamma_{\rm ex}$ is the exchange coupling, $\Gamma_0=\sigma_0\otimes\tau_3$, and $\Gamma_{1(2)}=\sigma_{1(2)}\otimes\tau_1$ with $\sigma$ and $\tau$ the Pauli matrices in the spin and orbital spaces. Equation~\eqref{eq:hl} includes a bias term $\delta\tau_3$ accounting for the asymmetric exchange coupling for the $p$-orbitals of the Bi and Te atoms~\cite{J.Wang_PRL_2019}. The values of $\delta$, $M_0$, $A_{1(2)}$, $B_{1(2)}$ $C$, $D_{1(2)}$ are specified in the supplemental materials (SM)~\cite{SM}. In the following, we will restrict to the low-temperature regime such that $f(\bm{k})\approx1$ for $\epsilon(\bm{k})<\epsilon_F$ and $f(\bm{k})\approx 0$ otherwise.

Without loss of generality, we set $\bm{E}=E\hat{\bm{x}}$ so the ${\rm{Tr}}(\Omega_{\nu\mu}^{km^j})$ tensor reduces to a vector, whose direction corresponds to that of $\bm{H}^T_j$ according to Eq.~\eqref{eq:SOT}. Numerically, we find that $\bm{H}^T_j$ is in the $x$ direction (collinear with $\bm{E}$) with its amplitude maximized on the outermost SLs. Since there is no consensus on the value of $\gamma_{\rm ex}$, we plot $H_{1,x}^{\rm T}$ (acting on the bottom SL) as a function of $\gamma_{\rm ex}$ for the AFM configuration, $\bm{m}^j=(-1)^{j+1}\hat{\bm{z}}$, from 2 to 6 SL thick. Inner SLs are subject to substantially weaker SOTs than the outermost SLs (but the dependence of $H_{j,x}^{\rm T}$ on $\gamma_{\rm ex}$ is similar in all SLs)~\cite{SM}. We have excluded the 1-SL case where the SOT vanishes identically (as it is prohibited by the inversion symmetry in linear response~\cite{singleSL}). In the odd-SL cases $H_{1,x}^{\rm T}$ changes non-monotonically with a sharp turn~\cite{Diverge}; whereas in the even-SL cases it varies monotonically. Figure~\ref{fig:SOT&band}(b) plots the $\Gamma$-point band gap versus $\gamma_{\rm ex}$, where gap closing appears at the sharp turn of $H_{1,x}^{\rm T}$ for each odd-SL case; the total Chern number $C_{tot}$ (including all bands below $\varepsilon_F$) transitions from $0$ to $-1$ across this point. The even-SL cases are supposed to be axion insulators, where the band topology is characterized by the axion field rather than $C_{tot}$~\cite{AxionInsulator,LayerChernNumber}.

\begin{figure}[t]
\includegraphics[width=\linewidth]{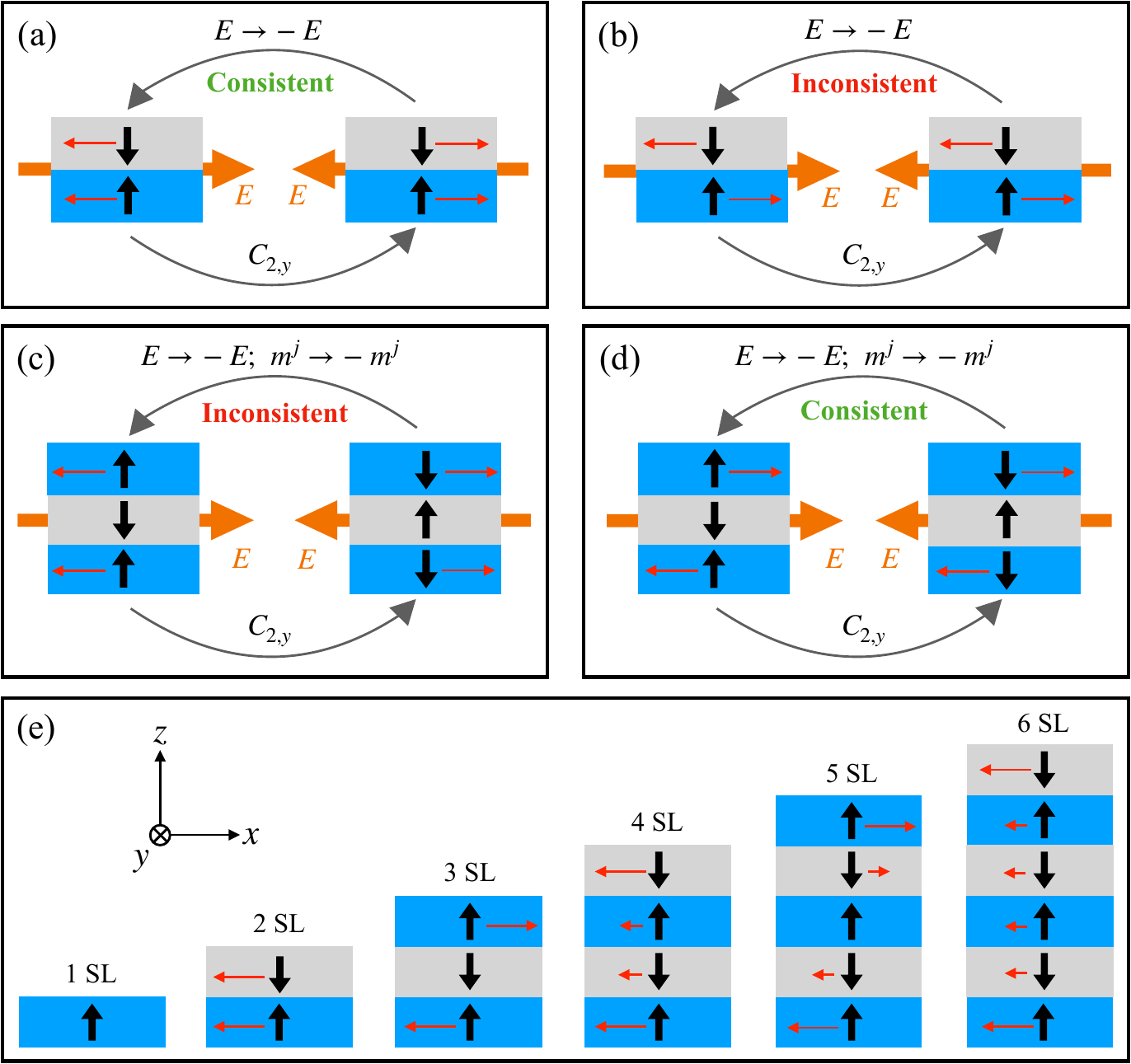}
\caption{(a)-(d) Symmetry analysis of the SL-resolved SOT fields $H^{\rm T}_{j,x}$ (red arrows) for 2-SL and 3-SL $\rm MnBi_2Te_4$. (e) The pattern of SOT fields from 1 SL to 6 SL. The rotation axis for $C_{2,y}$ locates at the geometric center of the sample.}
\label{fig:symmetry} 
\end{figure}

Comparing Fig.~\ref{fig:SOT&band}(a) with (b) implies that the SOT fields are subtly correlated to but not dictated by the band topology. For an odd-SL $\rm MnBi_2Te_4$, the SOT does not vanish in the $C_{tot}=0$ phase (normal insulator) although it is notably stronger in the $C_{tot}=-1$ phase (QAH insulator). In the even-SL cases (axion insulators), the SOT is strong despite that $C_{tot}=0$ throughout the whole plot range. To demystify these striking properties, we resort to the \textit{layer-resolved Chern number} defined specifically for each SL~\cite{LayerChernNumber,MBTTunnelJunction}. Figure~\ref{fig:SOT&band}(c) and (d) plot the amplitudes of the bottom-SL Chern number $C_1$ together with $H^{\rm T}_{1,x}$ as functions of $\gamma_{\rm ex}$ for the 3-SL and 4-SL cases, respectively. The insets further elaborate the SL-resolved Chern numbers $\{C_j\}$ and the SL-resolved SOT fields $\{H^{\rm T}_{j,x}\}$ for all $j$ involved. More details about other cases are left to the SM~\cite{SM}. We observe three crucial features from these results. First, the SOT acting on an outermost SL strongly correlates with the Chern number associated with that SL. Second, $C_j$ distributes symmetrically among the SLs while $H^{\rm T}_{j,x}$ distributes antisymmetrically in an odd-SL case; they swap their symmetry patterns in an even-SL case. Third, while $C_{tot}=\sum_jC_j$ delineates the band topology for the two phases in the odd-SL cases, it does not speak for the SOT. For example, $C_{tot}=0$ in the normal insulator phase only because $C_1=C_3=-C_2/2$, while the SOT fields satisfy $H^{\rm T}_{1,x}=-H^{\rm T}_{3,x}$. Here, $H^{\rm T}_{2,x}$ in the middle SL vanishes owing to the antisymmetric SL distribution (which will soon be explained in Fig.~\ref{fig:symmetry}) even though $C_2$ is finite.

We emphasize that the SOT originates from the Berry curvature $\Omega^{km}$ lying in the mixed momentum-magnetization space, whereas the band topology is determined by $\Omega^{kk}$ residing only in the momentum subspace. These two quantities satisfy an effective Faraday's relation~\cite{Bianchi} but they do not reflect each other explicitly. That explains why the SOT cannot be deduced directly from the band topology as discussed earlier. One could of course define a topological number for $\Omega^{km}$ by integrating it over the mixed $m$-$k$ space~\cite{NC_Hanke2017mixed}, but this quantity does not determine the SOT. What characterizes the SOT strength is the \textit{torkance}---the SOT field over the applied electric field [see Eq.~\eqref{eq:SOT}], which is of order $10^{-4}\rm{Oe\cdot m/V}$ in the outermost SL. This is three orders of magnitude larger than the topological magnetoelectric effect enabled by the axion field in $\rm MnBi_2Te_4$~\cite{SM} and one order of magnitude larger than the SOT in mixed Weyl semimetals~\cite{NC_Hanke2017mixed}.

We can understand the SL-resolved patterns of the SOT fields by analyzing the symmetry of the perturbed states (\textit{i.e.}, in the presence of $\bm{E}=E\hat{\bm{x}}$). As illustrated in Fig.~\ref{fig:symmetry}(a), a two-fold rotation around the $y$ axis ($C_{2,y}$) in a 2-SL sample amounts to flipping the direction of $\bm{E}$ while leaving the magnetic configuration and the atomic structure unchanged, which should lead to a sign change of $\bm{H}_j^{\rm T}$. Therefore, the system returns to itself under the combined operation of $C_{2,y}$ and $\bm{E}\rightarrow-\bm{E}$. We show comparatively in Fig.~\ref{fig:symmetry}(b) why opposite SOT fields $H_{1,x}^{\rm T}=-H_{2,x}^{\rm T}$ would lead to inconsistencies with linear response (\textit{i.e.}, $\bm{H}_j^{\rm T}$ flips sign as $\bm{E}$ reverses). Similarly, a $C_{2,y}$ operation on a 3-SL sample flips not only the $\bm{E}$ field but also the magnetization in all SLs, hence an invariant transformation involves $C_{2,y}$, $\bm{E}\rightarrow-\bm{E}$, and $\bm{m}^j\rightarrow-\bm{m}^j$, as demonstrated in Fig.~\ref{fig:symmetry}(c) and~(d). Consequently, the 3-SL case is compatible with linear response only when $H_{j,x}^{\rm T}$ is antisymmetric among the constituent SLs. These symmetry arguments can be generalized into more SLs, as schematically shown in Fig.~\ref{fig:symmetry}(e). The red arrows indicating $\bm{H}_j^{\rm T}$ are exaggerated for the inner SLs to improve visual clarity; their exact magnitudes are shown in Fig.~S2~\cite{SM}.

\textit{Magnetic resonances.}---In light of the symmetry of SOT depicted in Fig.~\ref{fig:symmetry}, we study the magnetic resonances in the 2-SL and 3-SL cases to exemplify the even-odd contrast. A 2-SL $\rm MnBi_2Te_4$ can be modeled as a collinear two-sublattice antiferromagnet affording two chiral resonance modes of frequencies $\omega^r_{1(2)}=\sqrt{\omega_A (2\omega_E+\omega_A)}\pm \omega_0$~\cite{C.Kittel_Phys.Rev_1952}, where $\omega_E$ is the inter-SL Heisenberg exchange interaction expressed in angular frequency, $\omega_A$ is the perpendicular easy-axis anisotropy, and $\omega_0=\gamma H_0$ is the bias magnetic field in the $z$ direction. Because an in-plane ac electric field $\tilde{E}e^{\mathrm{i}\omega t}$ (phasor notations adopted hereafter) generates the same SOT field on each SL, the 2-SL AFM resonance induced by the SOT is physically similar to that driven by a microwave, the detailed discussion on which is left in the SM~\cite{SM}.

\begin{figure}[t]
\centering
\includegraphics[width=\linewidth]{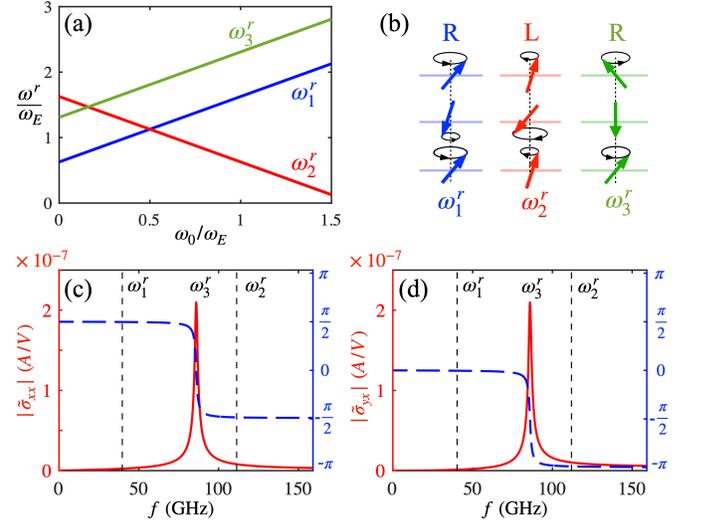}
\caption{(a) Resonance frequencies of a 3-SL $\rm MnBi_2Te_4$ varying with a perpendicular magnetic field (scaled into $\omega_0=\gamma H_0$). (b) An illustration of SL-specific magnetic precessions in each eigenmode. (c) $\tilde{\sigma}_{xx}$, and (d) $\tilde{\sigma}_{yx}$, plotted for their amplitudes (solid red) and phases (dashed blue) as functions of the driving frequency $f=\omega/2\pi$ for a 3-SL $\rm MnBi_2Te_4$. Parameters: $\alpha_{\rm G}=0.01$ (damping), $\gamma_{\rm ex}=75\ \rm meV$, $\hbar \omega_E=0.272\ \rm meV$ and $\hbar \omega_A=0.084\ \rm meV$~\cite{S.Q.Yang_PRX_2021}.}
\label{fig:AEmode}
\end{figure}

The 3-SL case, however, is quite non-trivial. Solving the coupled LLG equations for the SL-specific magnetization gives three distinct resonance modes
\begin{subequations}
\begin{align}
\omega^r_1&=\sqrt{\omega_A^2+3\omega_A\omega_E+\omega_E^2/4}-\omega_E/2+\omega_0 \label{eq:eigen-freqa},\\
\omega^r_2&=\sqrt{\omega_A^2+3\omega_A\omega_E+\omega_E^2/4}+\omega_E/2-\omega_0 \label{eq:eigen-freqb},\\
\omega^r_3&=\omega_A+\omega_E+\omega_0
\label{eq:eigen-freqc},
\end{align}
\end{subequations}
which are plotted in Fig.~\ref{fig:AEmode}(a). The SL-specific motions for each mode are illustrated in Fig.~\ref{fig:AEmode}(b). The right-handed mode $\omega_1^r$ (blue) and the left-handed mode $\omega_2^r$ (red), while having counterparts in the 2-SL case, are non-degenerate at zero field, which can be attributed to the uncompensated magnetization in the ground state of 3 SL $\rm MnBi_2Te_4$. In addition, we identify an exotic right-handed mode $\omega_3^r$ (green) in which the top and bottom SLs precess out-of-phase while the middle SL stays stationary as the instantaneous exchange torque exerting on it by the neighboring SLs exactly cancel. The top and bottom SLs do not directly couple; they affect each other indirectly through the middle SL.

Leveraging the $\omega_3^r$ mode calls for a staggered ac field acting oppositely on the top and bottom SLs while leaving the middle SL unperturbed, which coincides with the SL-contrasting SOT field depicted in Fig.~\ref{fig:symmetry}(d). Therefore, by virtue of the SOT, the $\tilde{E}e^{\mathrm{i}\omega t}$ field is able to induce the resonance of the $\omega_3^r$ mode. We numerically confirmed that the symmetry of SOT shown in Fig.~\ref{fig:symmetry} persists for fairly large-angle precessions and the SOT fields remain almost independent of the direction of $\bm{m}^j$ up to about $\theta=\pi/3$ (see Fig.~S9~\cite{SM}). Note that the $\omega_3^r$ mode is blind to microwaves because at the resonance frequency the wavelength far exceeds the SL spacing (so the oscillating magnetic field is SL independent). Moreover, while the Oersted field arising from $\tilde{E}e^{\mathrm{i}\omega t}$ is also staggered, it is much weaker than the SOT field~\cite{SM}. Consequently, observing the resonance of the $\omega_3^r$ mode provides an unequivocal way to verify the SOT.

By including the ac SOT field into the linearized LLG equations using phasor notations, we can solve the dynamical susceptibility as~\cite{SM}: $\tilde{\chi}_\parallel\equiv\tilde{m}^1_x/\gamma\tilde{H}^{\rm T}_{1,x}=\tilde{m}^3_x/\gamma\tilde{H}^{\rm T}_{3,x}=- (\mathrm{i}\alpha_{\rm G}\omega+\omega_3^r)/[\omega^2-(\mathrm{i}\alpha_{\rm G}\omega+\omega_3^r)^2] $ and $\tilde{\chi}_\perp\equiv\tilde{m}^1_y/\gamma\tilde{H}^{\rm T}_{1,x}=\tilde{m}^3_y/\gamma\tilde{H}^{\rm T}_{3,x}=\mathrm{i}\omega/[\omega^2-(\mathrm{i}\alpha_{\rm G}\omega+\omega_3^r)^2]$. The dynamical susceptibility is hard to measure directly because the oscillating SOT field is an intermediate quantity generated by $\tilde{E}e^{\mathrm{i}\omega t}$---the true driving force. To detect the SOT-induced resonance electronically, we should also consider the inverse effect of the SOT.

\textit{Adiabatic charge pumping.}---As the SL-dependent magnetization is driven into motion, the precessing magnetic moments will in turn generate a pure adiabatic current according to Eq.~\eqref{eq:pumping}~\cite{Thouless1983,QuantumPumpNiu,Ueda_PRB_2012}. In contrast to transport currents, an adiabatic current is not accompanied by Joule heating (\textit{i.e.}, it is dissipationless) and it decays rapidly when the system goes off-resonance~\cite{Oppen_PRL_2013,arrachea2015nanomagnet,J.Tang_PRB_2022}. Therefore, in our pure voltage-driven system, the pumped adiabatic current $\bm{J}^{\rm p}$ directly signals the onset of magnetic resonances. The overall effect combining the SOT, LLG equations, and adiabatic charge pumping manifests as a linear response relation: $\tilde{J}_\mu^{\rm P}(\omega)=\tilde{\sigma}_{\mu\nu}(\omega)\tilde{E}_{\nu}(\omega)$. For the 3-SL case, by eliminating the magnetic degrees of freedom, we obtain the ac conductivity as
\begin{subequations}
\begin{align}
    \tilde{\sigma}_{xx}(\omega)&=\mathrm{i}\omega\tilde{\chi}_\parallel(\omega)\frac{2e^2a_0^2}{\hbar S}\left\langle{\rm Tr}(\Omega_{xx}^{km^1})\right\rangle^2, \label{eq:sigmaxx}\\
     \tilde{\sigma}_{yx}(\omega)&=\mathrm{i}\omega\tilde{\chi}_\perp(\omega)\frac{2e^2a_0^2}{\hbar S}\left\langle{\rm Tr}(\Omega_{yy}^{km^1})\right\rangle^2,\label{eq:sigmayx}
\end{align}
\end{subequations}
where $\langle\cdots\rangle=1/(2\pi)^2\int {\rm d}^2kf(\bm{k})(\cdots)$ denotes the average over the first Brillouin zone~\cite{SM}. We plot the amplitude and phase of $\tilde{\sigma}_{xx}$ and $\tilde{\sigma}_{yx}$ as functions of the driving frequency $f=\omega/2\pi$ in Fig.~\ref{fig:AEmode}(c) and~(d), respectively. Remarkably, the amplitude $|\tilde{\chi}_{\parallel(\perp)}(\omega)|$ only has a single peak at $\omega_3^r$, indicating that the SOT exclusively excites the $\omega_3^r$ mode; it does not drive the $\omega_1^r$ and $\omega_2^r$ modes at all~\cite{QAH_and_pumping}. This confirms our expectation based on the symmetry of SOT. By contrast, a microwave source can only drive $\omega_1^r$ and $\omega_2^r$ but not $\omega_3^r$ (see Fig.~S8). Overall, the system behaves as an insulator off resonance while it admits pure adiabatic current on resonance. Similar to the SOT, the pumped current is significantly suppressed in the topological trivial regime.

\textit{Mechanical efficiency.}---The phase of $\tilde{\sigma}_{xx}$ varying over $\omega$ has profound physical implications. For instance, the electric response of a 3-SL sample [shown in Fig.~\ref{fig:AEmode}(c)] turns from capacitance-like into inductance-like as $\omega$ crosses $\omega_3^r$, resembling the behavior of a parallel LC-resonance~\cite{SM}. By acquiring an emergent reactance (capacitance and inductance) originating from the combined action of the SOT and its reciprocal effect, the system can function as an adiabatic quantum motor bearing zero Ohm's conduction, thus converting all input electric power into magnetic dynamics~\cite{Oppen_PRL_2013,arrachea2015nanomagnet,J.Tang_PRB_2022} without loss. In other words, energy is consumed only by the Gilbert damping but not through Joule heating. To confirm this remarkable property, we benchmark the time-averaged power consumption of the magnetic dynamics $P_M=\alpha_{\rm G}\hbar S(l_xl_y/a_0^2)\sum_j\overline{|\dot{\bm{m}}^j|^2}$ against the average input electric power $P_J\equiv\overline{J^{\rm p}_x E_x}l_xl_y$ (with $l_x$ and $l_y$ labeling the lateral dimensions), which gives rise to a mechanical efficiency~\cite{SM}
\begin{align}
     \xi=P_M/P_J=\alpha_{\rm G}\omega \left(|\tilde{\chi}_\parallel|^2+|\tilde{\chi}_\perp|^2\right)/|{\rm Im}(\tilde{\chi}_\parallel)|=1, \label{eq:efficiency}
\end{align}
in the absence of leakage currents and other imperfections. As a comparison, $\xi$ is only on the order of $1\%$ in current-driven 3D heterostructures where Joule heating dissipates most of the input electric power~\cite{J.Tang_PRB_2022}.

\begin{acknowledgments}
This work is supported by the Air Force Office of Scientific Research under Grant No. FA9550-19-1-0307. We sincerely thank F. Xue, H. Zhang, B. Lian, E. Del Barco, A. Kent, Y.-H. Li, S. Singh and Q. Niu for fruitful discussions.
\end{acknowledgments}

\bibliography{ref}

\begin{thebibliography}{62}
\bibitem [{Ons()}]{Onsager}%
  \BibitemOpen
  \href@noop {} {}\bibinfo {note} {{The sum over band index is omitted in
  Eq.~(3) for succinctness, but it has been taken care of in our calculations.
  It is not a coincidence that the SOT and adibatic charge pumping depend on
  the reciprocal off-diagonal components of the Berry curvature tensor. This is
  actually a footprint of the Onsager reciprocal relations, fulfilling
  thermodynamic requirements.}}\BibitemShut {Stop}%
\bibitem [{Div()}]{Diverge}%
  \BibitemOpen
  \href@noop {} {}\bibinfo {note} {{These sharp turns should actually be
  discontinuous jumps because of band closing at the transition points where
  the adiabatic condition is violated. The continuity of the SOT across the
  phase transitions is attributed to numerical inaccuracies of parameter
  sampling near the diverging points of the integrands.}}\BibitemShut {Stop}%
\bibitem [{QAH()}]{QAH_and_pumping}%
  \BibitemOpen
  \href@noop {} {}\bibinfo {note} {{The peak values of $\tilde{\sigma}_{xx}$
  and $\tilde{\sigma}_{yx}$ are approximately two orders of magnitude smaller
  than the QAH conductivity $e^2/h$. Consequently, $J_y^{\rm p}$ could be
  overwhelmed by the QAH current. Nonetheless, the phase of
  $\tilde{\sigma}_{yx}$ is $\pi/2$ behind $\tilde{E}_x$ on resonance, while the
  QAH signal is in phase with $\tilde{E}_x$, so the two effects can be
  distinguished. After all, $J_x^{\rm p}$ is unambiguous as it is unique to the
  topological charge pumping.}}\BibitemShut {Stop}%
\end{thebibliography}

\end{document}